\documentclass[journal]{IEEEtran}
\usepackage{mathtools}
\usepackage{amsmath,eqnarray}
\usepackage{amssymb}
\usepackage{textcomp}
\usepackage{graphics}
\usepackage{epstopdf}
\usepackage{cite}
\usepackage{float}
\usepackage{algorithm,algorithmic}
\usepackage[left=2.5cm,right=2.5cm,top=2.5cm,bottom=2.5cm]{geometry}
\usepackage{dsfont} 
\DeclareMathOperator*{\argmax}{arg\,max}
\begin{document}

\title{How Consumer Empathy Assist Power Grid in Demand Response}

\author{Mojtaba~Abolfazli,~\IEEEmembership{Student Member,~IEEE,}
	June~Zhang,~\IEEEmembership{Member,~IEEE,}
	and~Anthony~Kuh,~\IEEEmembership{Fellow,~IEEE}\\
	Department of Electrical Engineering, University of Hawaii at Manoa, Honolulu, HI 96822\\ Email: \{mojtaba, zjz, kuh\}@hawaii.edu}

%

\maketitle

\begin{abstract}
This paper investigates interaction among residential electricity users and utility company in a distribution network with the capability of two-way communication provided by smart grid. The energy consumption scheduling of electricity users is formulated as a game-theoretic problem within a group where all players are not totally selfish. Considering altruistic behavior of human decision-making, altruistic player action to other players actions can be influenced by recognizing the well being of others. The proposed model captures the empathy of electricity users in energy consumption scheduling and how this behavior affect peak demand and electricity prices. Numerical results demonstrate that both residential users and utility company can benefit through the channel of empathy.
\end{abstract}

\IEEEpeerreviewmaketitle

\section{Introduction}
Demand Side Management (DSM) commonly refers to programs implemented by utility companies to control the energy consumption at the customer side of the meter. These programs help to utilize available facilities more efficiently without need for upgrading and installing new generation and transmission infrastructure. The power system usually face generation shortage in peak hours, which necessitates to shed part of consumers’ demand in order to maintain balance between generation and demand which is the key factor in network security. This problem is more challenging nowadays as power systems face more uncertainties and risks due to high penetration of renewable resources and more fluctuated consumption patterns with the introduction of Plug-in Electric Vehicles (PEVs). The active participation of consumers in Demand Response (DR) programs, which is triggered by incentive signals, is on of the best solutions to deal effectively with this problem \cite{qdr2006benefits}.\\
\indent
Game theory as the study of conflicts and cooperation among intelligent rational decision-makers is one of the most useful tools for modeling the behavior of consumers in response to electricity prices. The performed studies in this field are mainly emphasized on modeling the interactions among a group of consumers in an area which is supplied by a single utility/multiple utilities. In such environment, each user tries to purchase larger amount of power at lower prices to achieve higher welfare and the goal of utility company is to achieve higher welfare by setting higher electricity prices. In such environment, it is assumed that enough generation resources are available in the grid to supply demand of the users while from the realistic perspective, sometimes the power grid faces lack of generation capacity because of technical and non-technical reasons.\\
\indent
Here, we want to introduce a new approach for modeling the complicated interactions between electricity users in a grid with limited generation resources. Through the channel of empathy, users may be concerned with their usage effect on other users. According to the performed studies in economics and behavioral psychology, individuals show some degree of empathy towards other population members \cite{andreoni1990impure,bolton1997rationality,charness2002understanding,batson2011altruism}. In other words, individuals try to imagine themselves in others’ shoes and consider social preferences which results in more cooperation among population. In this paradigm, the overall utility function of each player is a combination of their own utility function and other individuals utility function, which means that well-being of others directly affects well-being of altruistic individuals. In \cite{shim2012influence, eksin2016disease}, it was shown that taking protective measures by infected individuals could have a significant effect on disease control.\\
\indent
Electricity as a public good can be investigated in this context. An altruistic consumer cares about the welfare of others and decrease or shift his demand correspondingly if he knows that generation resources in the grid are not adequate. The idea is that residential users do not care much to fulfill their demand within a specific time period and have more flexible consumption pattern. It is worth mentioning that all players are not altruistic and some of them might be selfish which only focus on their payoffs.  Considering the presence of selfish users, we investigate the impact of empathy on energy saving and electricity prices and determine how the energy consumption of selfish users changes when the network is populated with altruistic users.\\
\indent
The remainder of this paper is organized as follows. An overview of related works in game-theoretic DR is given in Section II. In Section III, we formulate the energy scheduling among a group of residential users with selfish and altruistic behaviors and propose a distributed algorithm for solving the problem. Numerical results are provided in Section IV and conclusions are drawn in Section V.
\section{Related Work}
By the introduction of smart grid, the access to potential capabilities of DR programs becomes much more viable because of two-way exchange of information between users and utility company. In such environment, the utility and users can interact by exchange of supply and demand information in order to make mutually beneficial decisions \cite{deng2015survey}. Different approaches have been employed to study demand response in smart gird paradigm and game theory is one of the most promising approaches due to the dynamic nature of interactive decision-making between users and utility. In the following, we explore several studies related to game-theoretic DR within three different groups. However, first we introduce some terms which are used frequently hereafter:
\noindent
\begin{itemize}
	\item \textbf{Players} denote decision makers of the game such as electricity users and utility.
	\item \textbf{Payoffs} describe expected benefits that each player obtains which could be maximizing revenue for utility or minimizing billing cost for user.
	\item \textbf{Actions} are decisions made by each player like scheduling energy consumption for user and setting electricity prices for utility company.
\end{itemize}

\subsection{Non-cooperative DR}
Non-cooperative game is a type of game that predicts individual payoffs among players that have conflicting interests without any communication and cooperation \cite{saad2012game}. In \cite{mohsenian2010autonomous}, a non-cooperative game was employed to model interactions between consumers to decrease their billing cost through energy consumption scheduling game. A coupled-constraints energy consumption scheduling for residential consumers was assessed in \cite{deng2014residential}. The authors in \cite{ma2014distributed} proposed a distributed control strategy with real time pricing feedback to remove peak demand in a noncooperative game formulation.
\subsection{Cooperative DR}
The cooperative game investigates coalition that may form between players as a result of external enforcement for cooperation and communication. In this type of game, users decide to cooperate with each other through communication to improve their payoffs. In fact, the main challenge is how to incentivize users to get together for cooperation or punish them to prevent non-cooperative behavior. In \cite{ma2015cooperative}, a punishment mechanism was proposed to prevent selfish behaviors, which led to Pareto-optimal solution. A repeated game along with punishment mechanism detected by deviation from cooperative consumption were used to encourage cooperative behavior among building managers \cite{ma2014cooperative}. Enforcing  cooperation behavior with Extended Joint Action Learning method as decentralized reinforcement learning method was studied in \cite{hurtado2018enabling}. In this study, a joint reward function was introduced with the aim of finding actions which maximize this function within a multi agent system.
\subsection{Stackelberg DR}
The Stackelberg or leader-follower game models the hierarchical interactions between players where the leader moves first and the follower chooses his strategy in response to that. Here, the leader would be the utility company or aggregator which sets prices and the follower is the end-user which determines load profile based on electricity prices. In \cite{chen2012optimal,kilkki2015optimized,yu2016real,el2017managing,latifi2017fully,lu2018data}, Stackelberg game with one leader and N-followers was employed to model DR problem. More generalized form of this problem is to have multiple utilities which compete each other to maximize their payoff by setting electricity prices and each user manages electricity portfolio by choosing among them \cite{maharjan2013dependable,chai2014demand}.\\
\indent
In this paper, we model empathy in non-cooperative DR within a group of residential users and single utility company with limited generation resources and explore how it affects electricity prices and total demand. By employing dual decomposition method, we show that the social maximization problem can be solved in a distributed manner without need for access to the generation cost of utility company and payoff functions of users.

\section{Model Description}
\begin{figure}[t!]
	\centering
	\includegraphics[width=2.8in]{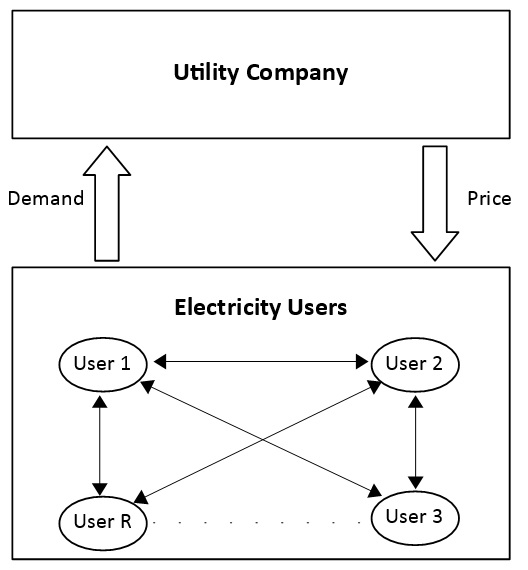}\\
	\caption{Interaction between consumers and utility company.}\label{Fig1.CU}
\end{figure}

The analytical description of the problem in a game framework is explained in the rest. It is assumed that there is a grid with multiple consumers and a single utility company that supplies demand. The consumers are communicating with each other and the utility company via the capability provided by smart grid as shown in Fig. \ref{Fig1.CU}.
\noindent

\subsection{Electricity Demand}
The set of residential consumers is denoted by \mbox{$\mathcal{R}$ $=\{1,2,\ldots,R \}$} and we assume that a day is divided into T time slots shown by \mbox{$\mathcal{T}$$=\{1,2,\ldots,T \}$}. The daily demand for consumer $r$ is denoted by \mbox{$\mathcal{D}_{r}$ $=\{x_{r,1},x_{r,2},\ldots,x_{r,T}\}$} and it is clear that the total hourly demand of users can be calculated by summation across $r$. Here we assume that each user has a minimum and maximum power consumption during each time slot:
\begin{equation}
x_{r,t}^{min}\leq x_{r,t} \leq x_{r,t}^{max},
\end{equation}
where the minimum value occurs for appliances which must be kept on within a specific time and level such as lighting and the maximum value shows highest possible value of demand at each time slot for each user.

\subsection{Utility Function}
The utility function shows the level of satisfaction obtained by each user as a function of energy consumption. Generally, quadratic form is used to describe the behavior of this function \cite{samadi2012advanced}:
\begin{equation}\label{Mark.Uti}
U(x)=
\begin{cases}
wx-(\alpha/2)x^2, & 0\leq x \leq w/\alpha \\
w^2/(2\alpha), & x > w/\alpha, \\
\end{cases}
\end{equation}

\noindent
where $x$ is the amount of energy consumption, $w$ is a positive parameter which varies for different users and $\alpha$ is a constant. The utility function has two main properties:

\noindent
\begin{itemize}
	\item It is an increasing function which is saturated as the user reaches the desirable consumption level which means that first order derivative is positive:
	\begin{equation}\label{Mark.Uti1d}
	\frac{\partial U(x)}{\partial x} >0, \hspace{0.4cm}  0\leq x <w/\alpha.
	\end{equation}
	\item It is a concave function which means that the marginal benefit to the user is decreasing as consumption increases:
	\begin{equation}\label{Mark.Uti2d}
	\frac{\partial^2 U(x)}{\partial x^2} <0, \hspace{0.4cm}  0\leq x <w/\alpha.
	\end{equation}
\end{itemize}

\begin{figure}[t!]
	\centering
	\includegraphics[scale=0.45]{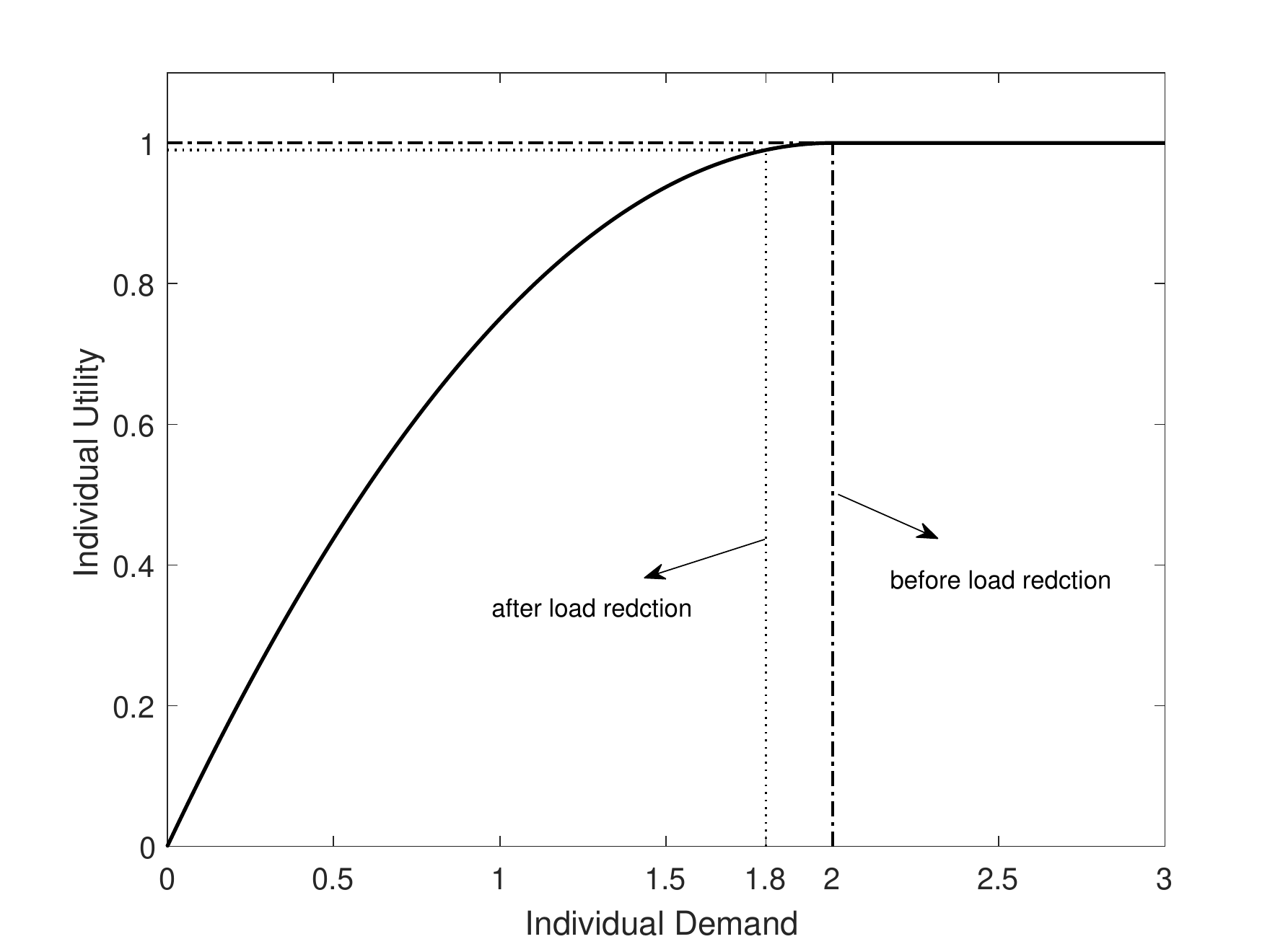}\\
	\caption{Impact of empathy on utility function of a single user.}\label{Fig2.UE}
\end{figure}

\indent
We assume that the consumers can be classified into two groups. The first group are selfish consumers where only concerned about their utility function. From the perspective of this group, other consumers are selfish as well. On the other side, there is another group  which concern about the others and try to consider the community preferences. For example, when the grid confronts generation shortage, two solutions are possible to keep the balance between load and generation. The first one is to shed demand which means some users will be disconnected. The second solution is that some users decrease their demand in that critical period to help the network operator and maybe themselves from getting disconnected. Traditionally, it can be obtained by offering monetary incentives for decrease in demand. However, in this paper, it is assumed that some users consider social welfare, motivated by different purposes such as environmental concerns or desire for equity, in energy consumption scheduling. This type of consumers called altruistic consider part of other users energy consumption as their own. In such situation, the altruistic consumers evaluate the condition and behave in such a way to prevent load shedding. We can model the behavior of this group as below:
\begin{equation}\label{Mark.NUti}
\widetilde{U}(x_{r})=U\bigg(x_{r} + \rho_{r} \bigg[x_r^{base}-\frac{\sum_{i\in \mathcal{R} \setminus \{r\} }x_{i}}{|\mathcal{R}|-1}\bigg]^+\bigg).
\end{equation}

\noindent
where $\rho_r$ is a coefficient that shows the level of empathy for user $r$, $x_r^{base}$ is the baseline of energy consumption for the user $r$ obtained from historical profile, $|\mathcal{R}|$ is the total number of users, and $[.]^+$ is the projection onto non-negative orthant. It should be noticed that the altruistic user compares its baseline consumption with the average of other participants, hereafter referred to as $\bar{x}_{-r}=\frac{\sum_{i\in \mathcal{R} \setminus \{r\} }x_{i}}{|\mathcal{R}|-1}$, and decides to decrease its demand if uses more than global average. Clearly, the value of $\rho_r$ for selfish users is zero. The maximum value of utility for a user with a non-zero empathetic parameter occurs at a level less than $x_r^{*}=w/\alpha$, as the optimum value obtained by \eqref{Mark.Uti}. This can be verified by maximizing \eqref{Mark.NUti} over $x$:

\begin{equation}\label{Mark.maxuti}
\begin{split}
x_r^*&=\argmax_{x_r} \hspace{0.1cm} \left[\frac{\partial \widetilde{U}(x_{r})}{\partial x_{r}}=0 \right]\\
&=\begin{cases}
w/\alpha- \rho_{r} (x_r^{base}-\bar{x}_{-r}), & x_r^{base} \geq \bar{x}_{-r}  \\
w/\alpha, & x_r^{base} < \bar{x}_{-r}. \\
\end{cases} \\
\end{split}
\end{equation}
\indent
It is clear that losing part of demand costs users and decrease their utility function. Due to the property of utility function, social welfare is higher when a group of people decrease part of their demand than the situation in which a few number of users get disconnected from the grid. To clarify this property, a sample utility function for a typical residential consumer is shown in Fig. \ref{Fig2.UE}. As it can be seen, the maximum utility happens at the demand of 2 kW. We can consider a network consisting of 10 identical residential users which faces generation shortage and can only supply 18 kW to the users. Since all consumers want to maximize their utility, we need 20 kW to fulfill their demand. In such condition, two different solutions are available to keep the balance between supply and demand as a critical requirement in maintaining power system stability \cite{kundur1994power}:
\begin{itemize}
	\item The first alternative is that the system operator disconnects one consumer from the grid. In such condition, the total utility function of consumers would be 9 units.
	\item The second alternative is that all of consumers decrease their demand by $10\%$ to compensate for the shortage. This means they use $1.8$ kW instead of $2$ kW. That causes the value of utility drops a little from $1$ to $0.99$. The aggregated utility in this situation is $10 \times (0.99)$ units which is $10\%$ higher than the former case.
\end{itemize}
A higher aggregated utility is attained without disconnecting any user from the grid by participation of users in demand reduction. In a game like this one, users may show altruism for getting benefits in long-term as the consequence of spreading social awareness.

\subsection{Cost Function}
The cost function is the cost of utility company for providing the electricity to end-users. In power systems, this function is usually increasing and strictly convex:

\noindent
\begin{equation}\label{Mark.CF1d}
\frac{\partial C(l)}{\partial l} >0,\hspace{0.25cm}  \frac{\partial^2 C(l)}{\partial l^2} >0.
\end{equation}

\noindent
The most common form of cost function is quadratic and is a function of total  available energy provided by the utility company \cite{wood2012power}. In this term, $a_t>0$, $b_t,c_t\geq 0$, and $l_t$ is total supplied demand and they can vary from one hour to another:

\noindent
\begin{equation}\label{Mark.CF}
C(l_{t})=a_t l_{t}^2+b_t l_{t}+c_t.
\end{equation}

\noindent
A simpler and more practical form of cost function is to assume that for each hour, the utility company sends users the price for one unit of electricity consumption to inform them about the electricity prices. We assume that maximum available generation for each time slot is bounded ($l_{t} \leq l_{t}^{max}$) as a result of limitation in resources or technical problems in the grid. The minimum available generation capacity set $l_{t}^{min}=\sum_{r\in \mathcal{R}} x_{r,t}$ to guarantee minimum required energy for each time slot.

\subsection{Objective Function}
From the global perspective, the objective is to increase the total utility of all the users in addition to decrease the expense of supplying energy by the utility company. In other words, the objective is to maximize social welfare (sum of utility function of all users minus the cost of supplying demand) by finding $x_{r,t}$ and $l_t$ in the following equation:

\noindent
\begin{align}\label{Mark.ObF}
\begin{split}
\textbf{maximize} \hspace{0.25cm} &\sum_{t\in \mathcal{T}} (\hspace{0.1cm} \sum_{r\in \mathcal{R}}\widetilde{U}(x_{r,t})-C( l_{t}))\\
\textbf{subject to} \hspace{0.25cm} &\sum_{r\in \mathcal{R}} x_{r,t} \leq l_{t}, \hspace{0.25cm} \forall t \in \mathcal{T} \\
&x_{r,t}^{min} \leq x_{r,t}\leq x_{r,t}^{max}\\
&l_{t}^{min} \leq l_{t} \leq l_{t}^{max},
\end{split}
\end{align}

\noindent
where $\widetilde{U}(x_{r,t})$ is defined in \eqref{Mark.NUti}, $l_{t}$ is the available energy in each hour which is limited in some critical time period such as peak hours, and $x_{r,t}^{min}$, $x_{r,t}^{max}$ denote constraints on minimum and maximum energy consumption in each hour for user $r$. It is clear that \eqref{Mark.ObF} is a convex function and has a maximum. \\
\indent
We can solve \eqref{Mark.ObF} for each time slot separately since this problem is not coupled in time. Therefore the Lagrangian form of \eqref{Mark.ObF} for a single time slot $t$ can be written as below:
\noindent
\begin{equation}\label{Mark.LF}
\begin{split}
\mathcal{L}(\textbf{x}_t,l_t,\lambda_t)= \sum_{r\in \mathcal{R}}\widetilde{U}(x_{r,t})-C(l_{t}) + \lambda_{t} ( l_{t} - \sum_{r\in \mathcal{R}} x_{r,t}),
\end{split}
\end{equation}
where $\textbf{x}_t=[x_{1,t},x_{2,t},\ldots,x_{R,t}]$ is the energy consumption vector and $\lambda_{t}$ is the Lagrangian multiplier at the time slot $t$.\\
\indent
The challenge with solving \eqref{Mark.LF} is that the utility company needs to know the utility function for each user to find the optimum solution. However, the utility function is private to the users. Therefore, instead of being able to solve the global optimum, we can only solve for the local optimum to each user. A good approach to solve this problem is to use Lagrangian dual problem in order to decouple the problem into separable sub-problems and use duality theory to show that local and global optimum are identical when duality gap is zero \cite{boyd2004convex}. The Lagrangian dual function is defined as the maximum value of \eqref{Mark.LF} over vector $\textbf{x}_t$:
\begin{align}\label{Mark.Ldual}
\begin{split}
\mathcal{D}(\lambda_t)&=\underset{\substack{x_{r,t}^{min}\leq x_{r,t} \leq x_{r,t}^{max} \\l_{t}^{min} \leq l_{t} \leq l_{t}^{max}}}{\textbf{maximize}} \hspace{0.1cm} \mathcal{L}(\textbf{x}_t,l_t,\lambda_t)\\
&= \sum_{r\in \mathcal{R}} \mathcal{D}_{r,t}^{S}(\lambda_t)+\mathcal{D}_{t}^{P}(\lambda_t),
\end{split}
\end{align}
where
\begin{equation}\label{Mark.duser}
\mathcal{D}_{r,t}^{S}(\lambda_t)=\underset{x_{r,t}^{min}\leq x_{r,t} \leq x_{r,t}^{max}}{\textbf{maximize}} \hspace{0.1cm} \widetilde{U}(x_{r,t})-\lambda_{t} x_{r,t},
\end{equation}
and
\begin{equation}\label{Mark.dutility}
\mathcal{D}_{t}^{P}(\lambda_t)=\underset{l_{t}^{min} \leq l_{t} \leq l_{t}^{max}}{\textbf{maximize}} \hspace{0.1cm} \lambda_{t} l_{t}-C(l_{t}).
\end{equation}

\indent
The dual problem is:
\begin{align}\label{Mark.dual}
\begin{split}
\underset{\lambda_t}{\textbf{minimize}} \hspace{0.25cm} &\mathcal{D}(\lambda_t)\\
\textbf{subject to} \hspace{0.25cm} &\lambda_t \geq 0.
\end{split}
\end{align}

\noindent
The first term in \eqref{Mark.Ldual} is dependent on users decisions and the second term is determined based on utility decision. Since the strong duality holds for a convex with affine constraints problem, the primal problem in \eqref{Mark.ObF} is equivalent to dual problem in \eqref{Mark.dual}, which means the optimality gap is zero and hence the solution of dual problem, $x_{r,t}^{*}$ and $l_{t}^{*}$, is the optimum solution in the primal problem \cite{boyd2004convex}.\\
\indent
The gradient projection method can be used to update Lagrangian multipliers as long as the dual function is differentiable \cite{wang2000convergence}. The Lagrangian multiplier $\lambda_t^{*}$ is calculated by the utility company in an iterative process and can be interpreted as the coordination signal, which is sent to the users for finding the optimum solution:

\begin{equation}\label{Mark.lambProj}
\lambda_t^{k+1}=[\lambda_t^{k}-\theta (l_{t}^{*} - \sum_{r\in \mathcal{R}} x_{r,t}^{*}(\lambda_t^{k}))]^{+},
\end{equation}

\noindent
where $\theta$ is a small positive step-size and $k$ denotes the iteration. Here we use local optimizer in \eqref{Mark.duser} for each user to find $x_{r,t}^{*}(\lambda_t^{k}$) and solve the dual problem iteratively.\\
\indent
The optimal value for each user depends on his consumption and also on how other users schedule their consumption. Here, it is assumed that each user announces his demand and has access to the demand of other users through the capability provided by smart grid \cite{mohsenian2010autonomous}. It is worth mentioning that the assumption for access to other users consumption does not put the privacy of users at risk since such data is required for meter reading and billing purposes. For finding the optimum solution, we need two levels of iterations. In the first level, the utility find optimum generation capacity from \eqref{Mark.dutility} and based on that and received consumption data from users, updates Lagrangian multipliers for sending to users. In the lower level, each user maximizes its payoff according to \eqref{Mark.duser}. Then, each user sends back its energy consumption value to the utility for updating generation capacity and Lagrangian multipliers. This process will continue until Lagrangian multiplier does not change. It is worth noting that altruistic users need average consumption of other users to decide whether or not weight it in their utility function as additional data. Since, users payoff function is strictly concave in terms of their own strategy set, it is resulted directly from \cite{rosen1965existence} that there is a unique Nash Equilibrium which is obtained by finding the optimum solution of the dual problem. \\
\indent
The distributed algorithm for finding the solution is summarized below. It must be executed for each time slot separately.

\begin{algorithm}
	\caption{for finding optimal solution}
	\begin{algorithmic}[1]
		\STATE \textbf{Initialization}\\
		\WHILE {Lagrangian multiplier changes}
		\FOR {each iteration $k$}
		\STATE Utility finds $l_t(\lambda_t^{k})$ based on \eqref{Mark.dutility} and received data from users
		\STATE Utility updates $\lambda_t^{k}$ based on \eqref{Mark.lambProj}
		\STATE Utility sends $\lambda_t^{k}$ and $\bar{x}_{-r,t}^{*}(\lambda_t^{k})$ to users (applicable for altruistic users)
		\STATE Each user computes the level of consumption $x_{r,t}^{*}(\lambda_t^{k})$ from \eqref{Mark.duser} 
		\STATE Each user broadcasts $x_{r,t}^{*}(\lambda_t^{k})$ to utility
		\STATE Set $ k \leftarrow k+1$
		\ENDFOR
		\ENDWHILE
		\STATE \textbf{return} $\lambda_t^{k*}$ and $x_{r,t}^{*}(\lambda_t^{k}) \hspace{0.1cm} \forall r \in \mathcal{R}$
	\end{algorithmic}
\end{algorithm}

\section{Numerical Results}
\begin{figure}[t!]
	\centering
	\includegraphics[scale=.6]{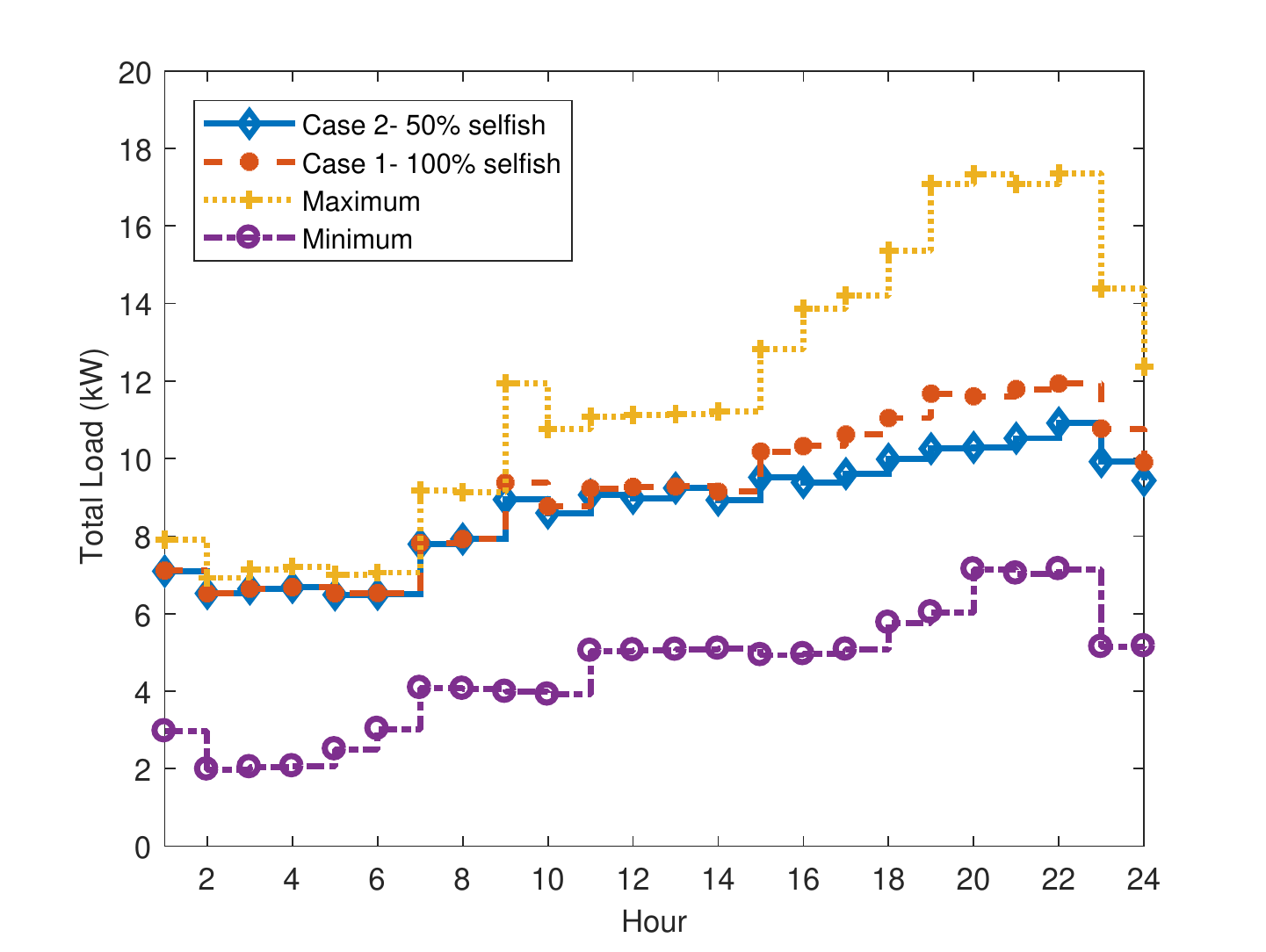}\\
	\caption{Aggregated energy consumption with/without empathy.}\label{Fig3.TD}
\end{figure}

\begin{figure}[t!]
	\centering
	\includegraphics[scale=.6]{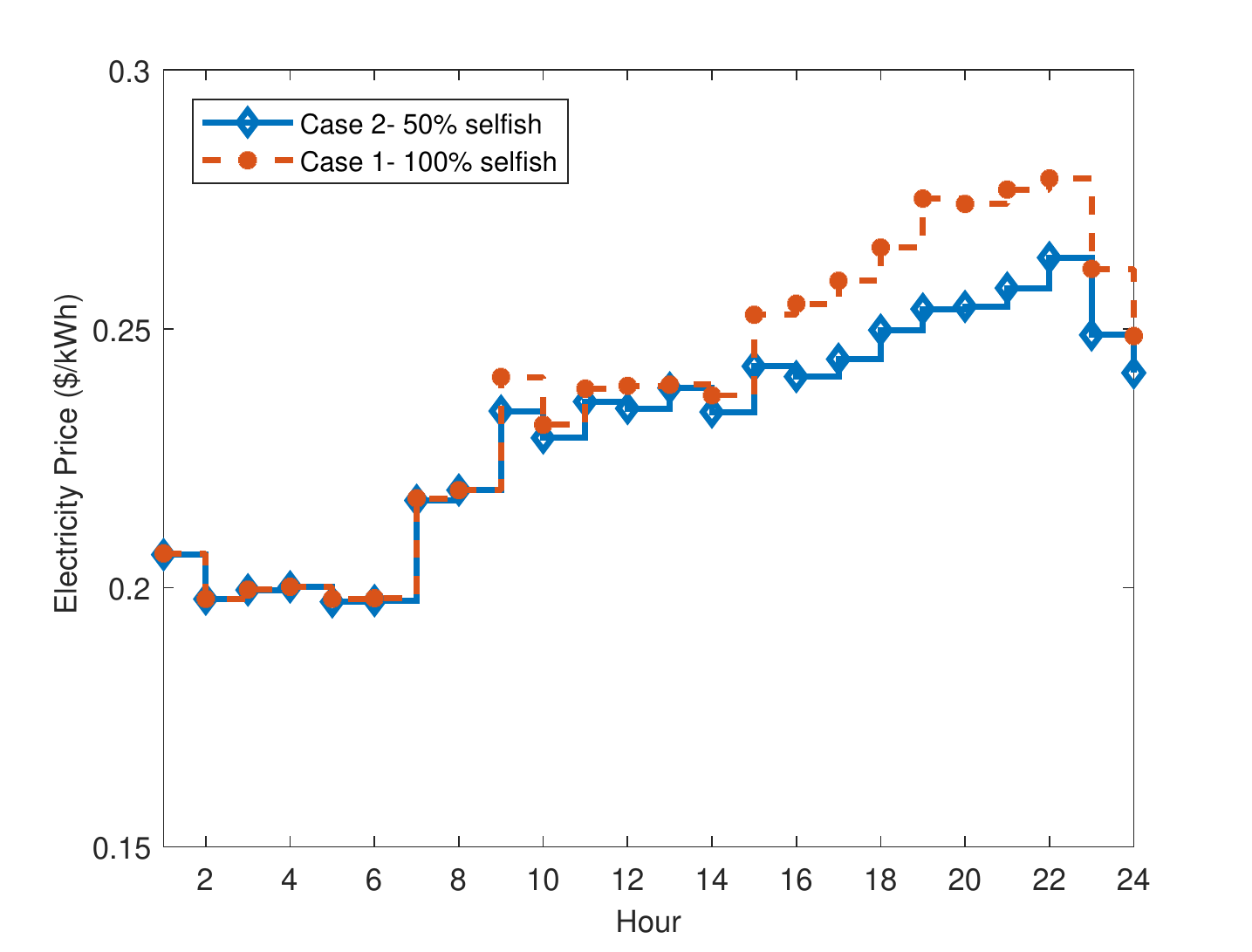}
	\caption{Electricity prices with/without empathy.}\label{Fig4.EP}
\end{figure}

In this section, the simulation results and performance of the model are investigated. In this model, a grid is composed of 10 residential consumers and a single utility company. The scheduling cycle is divided into 24 equal time slots in a typical day. The coefficients for the cost function of supplying demand by the utility company are $a_t=0.015$, $b_t=0.1$, and $c_t=0$ for all $t \in \mathcal{T}$. For the utility function of the users, $w$ is selected randomly from $[0.5,1.5]$ and $\alpha$ is fixed at $0.5$. For the level of empathy, we assume that altruistic users show a level of altruism towards others which are selected randomly from $[0.5, 1]$. The minimum and maximum generation capacity is set to the summation of minimum and maximum required energy by users for each time slot. Here, we consider two cases: 1) all users are selfish and 2) half of users are empathetic and the other half are selfish. Altruistic consumers show empathy if the average consumption of the population is less than their baseline consumption.\\
\indent
The total energy consumption patterns for the two cases are shown in Fig. \ref{Fig3.TD}. According to the results, both curves are within the permissible ranges, which is forced by subjecting users to the Lagrangian multiplier as the penalty cost of violating constraints. As it can be seen, total energy consumption decreases especially in peak hours as a result of altruistic behavior by some users. In fact, total amount of energy consumption decreases to $209.2$ kWh from $220.7$ kWh leading to a $5.2\%$ energy saving by the participation of altruistic users in DR. Obviously, it can assist the utility company to postpone the required investments for construction of new generation capacity and upgrading the network in the future. \\
\indent
Furthermore, by examining the simulation results, it is determined that the contribution of altruistic users decreases the electricity prices and consequently imposes less costs on the utility company for supplying users energy requirements. The hourly electricity prices curves for both cases are illustrated in Fig. \ref{Fig4.EP}. As a result of the linear relationship between marginal cost of electricity and total demand, higher load reduction implies higher decrease in electricity prices.

\begin{figure}[t!]
	\centering
	\includegraphics[scale=.6]{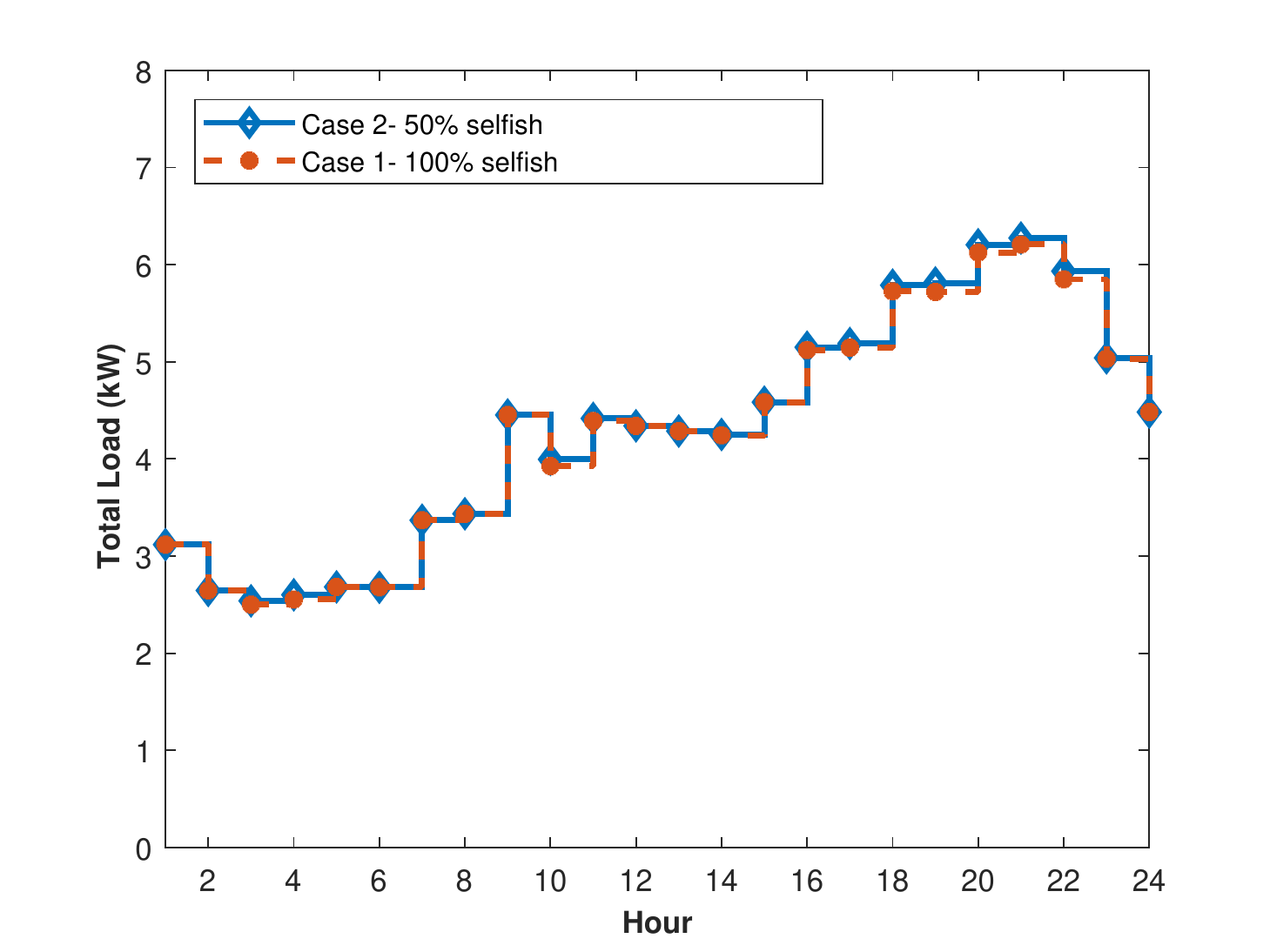}
	\caption{Aggregated energy consumption of same selfish users with/without empathy.}\label{Fig5.Selfish}
\end{figure}

In order to observe the effect of empathy on energy scheduling of selfish users, we plot the aggregated demand of selfish users with and without altruistic users in the grid. As shown in Fig. \ref{Fig5.Selfish}, the energy consumption of selfish users almost remains the same with a total increase of $0.5$ kWh during the time horizon, which is negligible comparing to $11.5$ kWh energy saving. In other words, selfish users could not make full use of energy saving by altruistic users.

\section{Conclusion}
In this paper, a new game-theoretic approach toward demand response by considering the altruistic behavior of residential users is examined. The dual decomposition method is employed to decompose the global problem into solvable local sub-problems which can be implemented in a decentralized manner. The simulation results show that both users and the utility company can benefit from this behavior by reduction in peak demand and electricity prices. The extension to this work can be studied by proposing a framework to motivate empathetic users for maintaining their altruistic behavior and spreading it to selfish users. 

\ifCLASSOPTIONcaptionsoff
  \newpage
\fi

\nocite{*}
\bibliography{IEEEabrv,reference}{}
\bibliographystyle{IEEEtran}

\end{document}